\documentclass{ws-procs9x6}

\begin{document}
\title{Black holes in the limiting curvature theory of gravity}

\author{Valeri P. Frolov$^*$}

\address{Theoretical Physics Institute, University of Alberta, \\
Edmonton, Alberta, Canada T6G 2E1\\
$^*$E-mail: vfrolov@ualberta.ca}

\begin{abstract}
We discuss a recently proposed limiting curvature theory of gravity and its application to the problem of singularities inside black holes. In this theory the growth of the curvature is suppressed by specially chosen inequality constraints included in the gravity action. We consider a case of a spherically symmetric four dimensional black hole and demonstrated that imposed curvature constraints modify a solution in the black hole interior. Instead of forming the curvature singularity the modified metric describes a space which is exponentially expanding in one direction and oscillating the other two directions. The spacetime is complete and its polynomial curvature invariants are uniformly bounded.
\end{abstract}


\bodymatter

\section{Introduction}\label{aba:sec1}

In this paper we discuss a recently proposed limiting curvature theory of gravity and its application to the problem of singularities inside black holes\footnote{
This is an extended version of the talk presented at the International Conference "Selected Topics in Mathematical Physics" Dedicated to 75-th Anniversary of I. V. Volovich. I know Igor Volovich for more than 50 years since 1967 when both of us were students of the Moscow State University. Our friendship with Igor played an important role in my life and I am grateful to him for this. Igor's scientific interests cover wide area of theoretical and mathematical physics. However, the subject of gravity and black holes was always one of his favorite, starting with our first joint publication \cite{Volovich:1976aw}.
}.
The main idea of this new approach is to modify the standard Einstein-Hilbert action by adding to it inequality constraints terms which control growth of the curvature. For properly chosen constraints solutions of the  equations satisfy the limiting curvature condition formulated by Markov \cite{Markov1,Markov2}. In a theory with inequality constraints
the additional terms in the modified action are chosen in such a form that they did not modify solutions of the gravity equations until a chosen curvature invariant reaches its critical value. Such subcritical solutions coincide with the solutions of the standard gravity equations. When the curvature constraint becomes saturated a regime of the solution changes to the supercritical one. After this transition one of the Lagrange multipliers accompanying the constraint, called a control function, becomes nonzero. In the supercritical regime the constraint equation is satisfied. This imposes an additional restriction on the metric. However,  the control function makes a modified set of equations consistent. It also provides the validity of the conservation law imposed by the covariance of the theory.

The theory of the limiting curvature gravity was developed in the following publications \cite{FZ1,FZ2,FZ3,FR}. They also contain many references to the papers where the material connected with the subject of the present talk can be found. In this talk I shall describe an application of the limiting curvature gravity theory to the problem of singularities inside black holes.

\section{A spherically symmetric black hole and its interior}

We consider a spherically symmetric vacuum black hole. In the subcritical regime, that is when the curvature is less than the critical one, its metric coincides with the Schwarzschild  solution of  Einstein equations
\begin{equation} \label{a1}
ds^2=-(1-\frac{2M}{r} )dt^2+\frac{dr^2}{1-\frac{2M}{r}}+r^2 d\omega^2\, .
\end{equation}
Here $d\omega^2=d\theta^2+\sin^2\!\theta\, d\phi^2$ is a metric on a two dimensional unit sphere.
Metric (\ref{a1}) is static and a corresponding Killing vector is $\boldsymbol{\xi}=\partial_{t}$.  This metric describes a black hole of mass $M$. The event horizon is located at $r=2M$. Inside the event horizon the metric (\ref{a1}) takes the form
\begin{equation}\label{a2}
ds^2=-\frac{dr^2}{f}+f dt^2+r^2 d\omega^2,\hspace{0.5cm} f=\frac{2M}{r}-1 .
\end{equation}
In the black hole's interior the metric function $f(r)$ is positive.
Kretschmann curvature invariant ${\cal K}=R_{\alpha\beta\gamma\delta}R^{\alpha\beta\gamma\delta}$
\begin{equation}
{\cal K}=\frac{48 M^2}{r^6} .
\end{equation}
grows  when $r$ decreases and it becomes infinite  at $r= 0$.

In what follows we assume that when the curvature reaches some limiting value the metric (\ref{a2}) will be modified. Later we describe a modified action in the adopted  limiting curvature model. At the moment we simply assume that this transition from subcritical to supercritical regime happens inside the black hole. We write the supercritical solution  in the form
\begin{equation}\begin{aligned}\label{metric}
ds^2=-b^2 d\tau^2+B^2 dt^2+a^2 d\omega^2 \, .
\end{aligned}\end{equation}
Here $a$, $b$ and $B$ are functions of  time $\tau$. Since this metric does not depend on the coordinate $t$ it has the same Killing vector $\boldsymbol{\xi}=\partial_{t}$ as the subcritical solution. Let us note that $\boldsymbol{\xi}^2>0$, so that $t$ is a spatial coordinate. One can rescale $t$ by multiplying it by a constant. This ambiguity is fixed by the condition that the norm of the Killing vector  is continuous at the transition point. There is another ambiguity in the metric (\ref{metric}). One can rescale its metric function $b(\tau)$. To fix it one can use the following  gauge-fixing condition $b(\tau)=1$.  For this synchronous gauge the coordinate $\tau$ coincides with the proper time. We shall use this gauge choice for study of  solutions of our model. However,  at the beginning it is useful to keep $b(\tau)$ arbitrary. The reason is the following. To simplify derivation of the equations we use a reduced action approach in which the action calculated for metrics of the form (\ref{metric}) becomes a functional depending on 3 functions $a(\tau)$, $b(\tau)$ and $B(\tau)$. Its variation with respect to these functions gives a set of equations which is equivalent to original complete set of equations restricted by the ansatz (\ref{metric}). This happens only if during the variation of the reduced action one keeps the function $b(\tau)$ arbitrary.

Let us note that the Schwarzschild metric (\ref{a2}) in the black hole interior can be presented in the form (\ref{metric}) by using the following relation between $r$ and $\tau$
\begin{equation}
\tau=\sqrt{r(2M-r)}+M \arcsin\Big(\frac{M-r}{M}\Big)+\frac{1}{2}\pi M .
\end{equation}
This relation shows that  $\tau$ coincides with a proper time along a  line with fixed values of $(t,\theta,\phi)$. The integration constant is chosen so that $\tau=0$ at $r=2M$. At $r=0$ one has $\tau=\pi M$.

\section{Curvature invariants}

In order to restrict growing of the curvature beyond a chosen limit one needs to specify these curvature invariants.
It is easy to check that the Riemann curvature tensor for the metric (\ref{metric}) has four non-vanishing components
\begin{eqnarray}\label{RIN1}
&& R_{\hat{\tau}\hat{t}\hat{\tau}\hat{t}}=-v,\hspace{0.5cm} R_{\hat{\tau}\hat{\theta}\hat{\tau}\hat{\theta}}=-q\, ,\\
&&R_{\hat{t}\hat{\theta}\hat{t}\hat{\theta}}=u, \hskip 0.77cm   R_{\hat{\theta}\hat{\phi}\hat{\theta}\hat{\phi}}=p .
\label{RIN2}
\end{eqnarray}
The hat over indices indicates that these components are calculated in the corresponding orthonormal tetrad frame.
The quantities $(p,q,u,v)$ which are eigenvalues of the Riemann tensor are
\begin{eqnarray}\label{pquv}
&&p=\frac{\dot{a}^2+b^2}{a^2b^2}, \hskip 0.7cm
q=\frac{\ddot{a}}{ab^2}-\frac{\dot{a}}{a}\frac{\dot{b}}{b^3},\\
&&u=\frac{\dot{a}\dot{B}}{a B b^2} , \hskip 1cm
v=\frac{\ddot{B}}{Bb^2}-\frac{\dot{B}}{B}\frac{\dot{b}}{b^3}  \label{uv}.
\end{eqnarray}
A dot in these expressions denotes a derivative with respect to $\tau$.

Let us consider a scalar function constructed from the Riemann curvature tensor. Relations (\ref{RIN1}) show that it can be presented as a function four variables  $(p,q,u,v)$. In particular, a scalar polynomial invariant  is a polynomial of $(p,q,u,v)$.
To restrict a value of such a polynomial it is sufficient to restrict the value of the curvature eigenvalues  $(p,q,u,v)$.
Let us also mention that the curvature eigenvalues $(p,q,u,v)$ can be  presented as functions of basic curvature invariants.
Namely,  the Riemann tensor can be decomposed into three irreducible terms: the Weyl tensor $C_{\alpha\beta\gamma\delta}$, the traceless Ricci tensor $S_{\alpha\beta}=R_{\alpha\beta}-\frac{1}{4}g_{\alpha\beta}R$, and the Ricci scalar $R$. Calculations give
\begin{equation}\begin{aligned}\label{RSS}
&{\cal C}^2\equiv C_{\alpha\beta\gamma\delta}C^{\alpha\beta\gamma\delta}=(p-q-u+v)^2\, ,\\
&R=2(p+2q+2u+v) ,\\
&{\cal S}^2\equiv S_{\mu\nu}S^{\mu\nu}=(p-v)^2+2(q-u)^2  ,\\
&{\cal S}^3\equiv S_{\mu}^{\nu}S_{\nu}^{\alpha}S_{\alpha}^{\mu}=-3(p-v)(q-u)^2 .
\end{aligned}\end{equation}
These equations are functionally independent and one can use them obtain  the eigen values $(p,q,u,v)$ in terms the above four basic invariants.

The curvature eigenvalues for the subcritical Schwarzschild metric are
\begin{equation} \label{pquvS}
v=p=-2q=-2u={2M\over r^3}\, .
\end{equation}

\section{Choice of inequality constraints}

For better understanding of the meaning of the curvature eigenvalues we consider two spacetimes of lower dimensions which are  slices of the spacetime (\ref{metric}).

{\bf 2D black hole section.} A two dimensional slice of (\ref{metric}) with  fixed angles $\theta$ and $\phi$ has the metric
\begin{equation} \label{gg}
d\gamma^2=-b^2(\tau)d\tau^2+B^2(\tau) dt^2 .
\end{equation}
2D curvature of this metric is ${}^{2\!}R=2 v$ , where $v$ is given in (\ref{pquv}). In the gauge $b=1$ one has
\begin{equation}
v={\ddot{B}\over B}={1\over 2}{}^{2\!}R\, .
\end{equation}
For the interior of the Schwarzschild black hole the metric function $B$ infinitely grows and reaches infinite value at finite proper time. As a result, the two-dimensional curvature infinitely grows. In order to avoid this, one should impose a restriction on the invariant $v$. A two-dimensional section of the supercritical solution with limiting value of $v=\Lambda$ is nothing, but a two-dimensional deSitter metric. Two-dimensional black holes with this type of a constraint were studied in detail in \cite{FZ1}. The main property of such a supercritical solution is that the deSitter like expansion of the black hole interior continues forever. In what follows we impose the following constraint condition
\begin{equation}\label{C1}
\Phi_1=v -\Lambda_1\le 0 \, .
\end{equation}

{\bf 3D  section.}
The second spacetime is a section $t=$const. It is three dimensional and has the metric
\begin{equation} \label{GG}
d\Gamma^2=-b^2(\tau)d\tau^2+a^2(\tau) d\omega^2 .
\end{equation}
The Einstein tensor for it is ${}^{(3)}G^{\mu}_{\nu}=-\mbox{diag}(p,q,q)$,
where $p$ and $q$ are given in (\ref{pquv}). This metric describes a $2+1$ closed homogeneous isotropic cosmology.
Relations (\ref{pquvS}) show that at the beginning of the supercritical phase this cosmology is contracting. For the Schwarzschild metric the scale parameter $a(\tau)$ reaches zero value at finite proper time. This is a Big Crunch where the invariants constructed from the three-dimensional Ricci tensor diverge. To escape this Big Cruch singularity it is sufficient to impose restrictions on the curvature eigenvalues $p$ and $q$. A similar problem for a "real" $3+1$ cosmological model was considered in \cite{FZ2}. One of the results of this study  is the following: to restrict curvature growth in such a model it is sufficient  to impose the following constraint condition
\begin{equation}\label{C2}
\Phi_2=p-\mu q -\Lambda_2 \le 0,\hspace{0.5cm} 0<\mu<1\, .
\end{equation}

By imposing constraints (\ref{C1}) and (\ref{C2}) one can restrict the values of the invariants $v$, $p$ and $q$. In such a model the value of the fourth invariant $u$ is also bounded. To demonstrate this we consider a supercritical solution where both constraints (\ref{C1}) and (\ref{C2}) are saturated. Condition $v=\Lambda_1$ implies that after leaving the subcritical Schwarzschild solution a positive function $V(\tau)$ continues to grow and
\begin{equation}
{\dot{V}\over V}=C \tanh( \sqrt{\Lambda_1}\tau +\tau_0)\, .
\end{equation}
Here $C$ is a positive constant and $\tau_0$ is a parameter defied by the initial conditions at the transition point. Thus
${\dot{V}/ V}<C$.  The invariant $p$ in the gauge $b=1$ is
\begin{equation}
p={\dot{a}^2+1 \over a^2} >\left({\dot{a}\over a}\right) ^2 \, .
\end{equation}
Hence, if this invariant is restricted, then the quantity $|\dot{a}/ a|$ is also restricted. But in the same gauge
\begin{equation}
u=\frac{\dot{a}}{a} \frac{\dot{B}}{B}\, .
\end{equation}
The above results imply that if the curvature eigenvalues $p$, $q$, and $v$ are bounded then the quantity
 $|u|$ is bounded as well.

\section{Reduced action with curvature constraint inequalities}

A starting point of the limiting curvature theory of gravity is an action
\begin{equation}\begin{aligned}\label{SSS}
S=\frac{1}{\kappa}\int d^4 x \,\sqrt{-g} L(g),\hspace{0.5cm}
\kappa=8\pi G ,
\end{aligned}\end{equation}
with the Lagrangian $L(g)$ of the form
\begin{equation}
L(g)=L_g +L_{c} .
\end{equation}
Here $L_g=\frac{1}{2}R$ is the Lagrangian for Einstein-Hilbert action and $L_{c}$ is a part of the Lagrangian responsible for inequality curvature constraints
\begin{equation} \label{LCC}
L_{c}=\sum_i \chi_i (\Phi_i +\zeta_i^2)\, .
\end{equation}
Here $\Phi_i$ are scalar functions depending on the curvature and a parameter $\Lambda_i$ which characterizes the value of the limiting  curvature.
The functions $ \chi_i$ and $\zeta_i$ are Lagrange multipliers associated with a constraint $i$. A variation with respect to  Lagrange multipliers gives the following equations
\begin{equation}
\Phi_i +\zeta_i^2=0,\hspace{0.5cm} \chi_i \zeta_i=0\, .
\end{equation}
The first of these equations has a solution only when $\Phi_i \le 0$. In the subcritical regime, that is when $\Phi_i < 0$ , it defines a non-zero value of $\zeta_i$. The second equation shows that in this regime the corresponding control function $\chi_i$ vanishes.
A transition from the subcritical to the supercritical regime happens when the corresponding curvature invariant reaches its critical value for which $\Phi_i = 0$. In this regime $\zeta_i=0$ and the control function $\chi_i$ does not vanish. Depending on the number and properties of the constraint functions $\Phi_i $ there may be several different supercritical regimes. A general evolution of the system with described inequality constraints can include both subcritical and different supercritical branches. A transition between these regimes is controlled by  functions $\chi_i$.

After these general remarks we return to our problem. As we already mentioned, a complete set of equations which determines a solution with metric  (\ref{metric}) can be obtained from the reduced action. The latter is just an action obtained from (\ref{SSS}) by substituting the ansatz (\ref{metric}) into it. Using symmetries of this metric one has
\begin{equation}\label{RA}
S=\frac{1}{\kappa}V {\cal S} ,\hspace{0.5cm} {\cal S}= \int d\tau\, a^2 b B \, {\cal L}(a,b,B) \, .
\end{equation}
Here
\begin{equation}
V=4\pi\int d t
\end{equation}
is (formally infinite) constant factor which does not affect the equations and can be omitted.
Hence,  we shall use the reduced action
\begin{eqnarray} \label{action}
& {\cal S}=\int d\tau a^2 b B \,{\cal L}\, ,\\
&{\cal L}={\cal L}_g+{\cal L}_c .\label{RedL}
\end{eqnarray}
Here ${\cal L}_g$ is the reduced form of the Einstein-Hilbert action
which after integration by parts is
\begin{equation}
{\cal S}_g=\int d\tau B \left[b-\frac{\dot{a}^2}{ b}-2\frac{a\dot{a} \dot{B}}{Bb}\right] .
\end{equation}
The Lagrangian for inequality constraints (\ref{LCC}) takes the form
\begin{equation}
{\cal L}_c=\sum_j \chi_j (\Phi_j+\zeta_j^2)\, .
\end{equation}

As we explained earlier to restrict curvature growing in our model it is sufficient to use two constrains. Thus we take
\begin{equation}
{\cal L}_c= \chi_1 (\Phi_1+\zeta_1^2)+\chi_2 (\Phi_2+\zeta_2^2)\, ,
\end{equation}
where the constraint functions $\Phi_1$ and $\Phi_2$ are defined by relations (\ref{C1}) and (\ref{C2}), respectively.

A complete set of equations which describes the evolution of the interior of a spherically symmetric black hole in the limiting curvature theory of gravity consists of
\begin{itemize}
\item Three gravitational equations obtained by the variation of the reduced action (\ref{RA}) over the metric functions $b(\tau)$, $B(\tau)$, and $a(\tau)$. Only two of these equations are independent. This property is a consequence of the covariance of the original action (\ref{SSS}).
\item Two equations obtained by variations over $\zeta_1$ and $\zeta_2$.
\item Two more constraint equations obtained by variations over $\chi_1$ and $\chi_2$.
\end{itemize}

\section{Results}

Solutions of the complete set of the equations have been studied in \cite{FZ3}. Here we reproduce main results of this analysis.

The subcritical Schwarzschild solution (\ref{a2}) is valid until  proper time $\tau=\tau_0$ at which one of the constraint functions $\Phi_i$ defined by (\ref{C1}) and (\ref{C2}) calculated for the curvature eigenvalues (\ref{pquvS}) reaches its critical value. It was shown that
\begin{itemize}
\item If the first constraint which becomes saturated is $\Phi_1$, then the transition point is regular and junction conditions uniquely determine the corresponding supercritical solution.
\item If $\lambda\equiv \Lambda_2/\Lambda_1> 1+\mu/2$ then the second constraint $\Phi_2$ is saturated at the proper time $\tau=\tau_1>\tau_0$. For $\lambda$ of order of one $\tau_1$ is close to $\tau_0$ and during the time between $\tau_0$ and $\tau_1$ the solution does not change significantly. If $\lambda=1+\mu/2$ the duration of this transition phase shrinks to zero. In such a case both constraints are switched simultaneously.
\item At the second phase where both constraints are saturated the corresponding supercritical solution can be found by solving the constraint equations\footnote{To simplify further equations we omit index 1, that is we put $\Lambda_1=\Lambda$.}
    \begin{equation}
    v=\Lambda,\hspace{0.5cm} p-\mu q=\lambda \Lambda\, .
    \end{equation}
\item At this stage $\zeta_1=\zeta_2=0$ and the gravitational equations determine evolution of the control functions $\chi_1$ and $\chi_2$. The analysis of the gravitational equations show that these control functions do not vanish, so that after the subcritical solution enters the supercritical regime it stays there forever.
\end{itemize}

Let us describe properties of this  supercritical solution. Constraints  $\Phi_1=0$ and $\Phi_2=0$ are decoupled equations. The first equation determines dependence of the metric function $B$ on $\tau$. Solving this equation one gets
\begin{equation}\begin{aligned}\label{BB2}
&B(\tau)=\frac{1}{2} \sqrt{3m^2-4} \cosh(\tau-\tau_0+\phi)\, ,\\
&\tanh\phi=\frac{1}{2}\frac{m}{\sqrt{m^2-1}}\, .
\end{aligned}\end{equation}
The constants in this solution are chosen so that it satisfies required continuity conditions at the transition point. A dimensionless parameter $m$ which enters this relation is defined as follows. Let us denote by $\ell$ a critical length determined by $\Lambda$, $\ell=1/\sqrt{\Lambda}$. Then
\begin{equation}
m=(2M/\ell)^{1/3}\, .
\end{equation}
As we already mentioned, in this regime the metric of a 2D black-hole slice of the 4D solution has constant 2D curvature and describes an expanding 2D deSitter universe.

Analysis of solutions of the second constraint $\Phi_2=0$ shows that
the metric function $a(\tau)$, which after the transition to the supercritical phase  decreases, at some moment of time $\tau$ reaches its minimal value $a\approx \ell$. After this $a(\tau)$ increases until it reaches its maximal value, which for $m\gg1$ is slightly  larger than $m\ell$. After this turning point $a(\tau)$ decreases again. In other words, the function $a(\tau)$ is periodic and the 3D slice with metric (\ref{GG}) describes an oscillating closed universe.

\section{Discussion}

\begin{figure}[!hbt]
    \centering
      \includegraphics[width=0.8\textwidth]{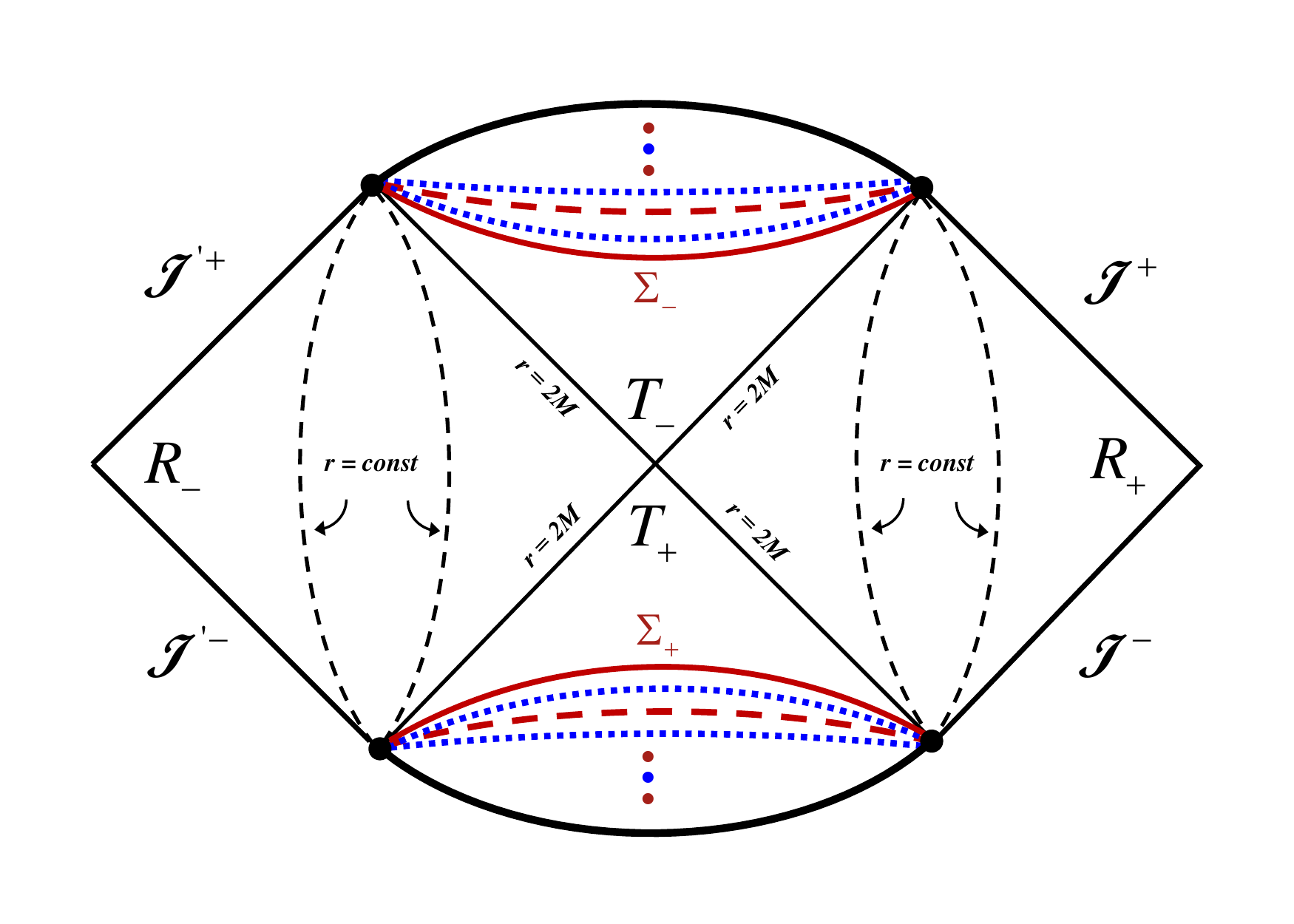}
    \caption{Conformal diagram for the eternal black hole in the limiting curvature model of gravity.}
    \label{Penrose}
\end{figure}

In the above discussion we focused on the structure of the spherically symmetric black hole interior in the limiting gravity model. We did not consider gravitational collapse of the matter which forms the black hole. In other words we study a solution which is usually called  an eternal black hole. This is maximally extended metric describing the gravitational field in vacuum. A conformal diagram for such an eternal black hole in our model is shown in Figure~\ref{Penrose}. This diagram contains four regions denoted by $R_{\pm}$ and $T_{\pm}$. Solid lines $\Sigma_-$ and $\Sigma_+$ separate subcritical and supercritical domains. The region below $\Sigma_-$ and above $\Sigma_+$ corresponds to a subcritical Schwarzschild-Kruskal solution. This region is nothing but a part of the Schwarzschild eternal black hole spacetime. It has two asymptotically flat spaces $R_{\pm}$. Lines $r=$ const in this regions are timelike. $T_-$ is the interior of the black hole. The diagram also contains white hole region $T_+$. At $\Sigma_{\pm}$ the curvature reaches its critical value, so that above $\Sigma_-$ and below $\Sigma_+$ the solution is supercritical. In the black hole interior the metric has 2D slice describing an expanding 2D deSitter universe. The scale factor $a(\tau)$ is oscillating there. Spatial slices of constant value of  $a$ (which are identical to slices of $\tau=$const) are shown by dotted lines inside $T_-$. Slightly above $\Sigma_-$ the value of $a$ decreases with time until it reaches its minimal value. This slice is shown by a dashed line inside $T_-$. After this $a(\tau)$ grows, reaches its maximal value, and then decreases again. With growing proper time these oscillations continue\footnote{In classification proposed in the paper \cite{Carballo-Rubio:2021wjq} the metric in the   black hole interior belongs to the class of the one-way (hidden) wormholes.}.

The metric in the white hole domain $T_+$ is similar with the only difference that a corresponding  2D deSitter  slice describes a contracting universe.  The spacetime of this eternal black hole is complete and the curvature invariants are uniformly bounded. In other words this model satisfies the limiting curvature condition.

There are several interesting questions connected with the results discussed in this paper. We made an assumption that the spacetime is spherically symmetric. Can small perturbations imposed on this metric become important? What happens if a black hole is rotating or it has an electric charge? Since the gravitational field in the black hole interior is strong one can expect intensive  particle creation in this domain. Can this effect change the described structure of the black hole interior? The Schwarzschild metric in the black hole interior near $r=0$ is similar to the Kasner metric with a special choice of its parameters. What are properties of a general type contracting anisotropic Kasner spacetime in the theory with limiting curvature? The last question is closely related to the following one. Can such a limiting curvature theory of gravity provide a natural mechanism of suppression of the anisotropy growing in the bouncing cosmologies? These problems are interesting and certainly require their further study.

\section*{Acknowledgments}

I am grateful to my collaborator Andrei Zelnikov for many fruitful discussions during our joint work.
The author thanks the Natural Sciences and Engineering Research Council of Canada and the Killam Trust for their financial support.



\bibliography{REF1}


\end{document}